\documentclass[superscriptaddress,tighten,twocolappendix,twocolumn]{openjournal}

\usepackage{grffile}             
\graphicspath {{Figures/}}
\usepackage[dvipsnames]{xcolor} 
\usepackage[bookmarks=false]{hyperref}
\hypersetup{
    colorlinks=true,
    linkcolor=blue,
    filecolor=magenta,      
    urlcolor=RedViolet,         
    citecolor=MidnightBlue,     
}
\usepackage[all]{nowidow}

\begin{document}
 
 \title{Interplanetary Dust Emission as a Foreground for the \emph{LiteBIRD} CMB Satellite Mission}
 
 \author{K.~Ganga} 
 \affiliation{Université de Paris, CNRS, Astroparticule et Cosmologie, F-75013 Paris, France}
 
 \author{M.~Maris}
 \affiliation{INAF/Trieste Astronomical Observatory, Via G.B. Tiepolo 11, ZIP I-34143 Trieste, Italy}
 \affiliation{Institute for Fundamental Physics of the Universe, Via Beirut, 2–4 I–34151, Grignano, Trieste, Italy}

 \author{M.~Remazeilles} 
 \affiliation{Instituto de Fisica de Cantabria (CSIC-Universidad de Cantabria), Avda. de los Castros s/n, E-39005 Santander, Spain}
 \affiliation{Jodrell Bank Centre for Astrophysics, Department of Physics and Astronomy, The University of Manchester, Manchester, M13 9PL, United Kingdom \\ \vspace{0.25cm} \\ for the \emph{LiteBIRD} collaboration \\ \vspace{0.15cm} \\ }
 
 \begin{abstract}
 
  As ever-more sensitive experiments are made in the quest for primordial CMB $B$ Modes, the number of potentially significant astrophysical contaminants becomes larger as well. Thermal emission from interplanetary dust, for example, has been detected by the \emph{Planck} satellite. 
  While the polarization fraction of this Zodiacal, or interplanetary dust emission (IPDE) is expected to be low, it is bright enough to be detected in total power. Here, estimates of the magnitude of the effect as it might be seen by the \emph{LiteBIRD} satellite are made. 
  The {\emph{COBE}} IPDE model from \cite{Kelsall1998} is combined with a model of the \emph{LiteBIRD} experiment's scanning strategy to estimate potential contamination of the CMB in both total power and in polarization power spectra.
  \emph{LiteBIRD} should detect IPDE in temperature across all of its bands, from 40 through 402\,GHz, and should improve limits on the polarization fraction of IPDE at the higher end of this frequency range. If the polarization fraction of IPDE is of order 1\%, the current limit from \emph{ISO}/CAM measurements in the mid-infrared, it may induce large-scale polarization $B$ Modes comparable to cosmological models with an $r$ of order $10^{-3}$. In this case, the polarized IPDE would also need to be modeled and removed.
  As a CMB foreground, IPDE will always be subdominant to Galactic emissions, though because it caused by emission from grains closer to us, it appears variable as the Earth travels around the Sun, and may thereby complicate the data analysis somewhat. But with an understanding of some of the symmetries of the emission and some flexibility in the data processing, it should not be the primary impediment to the CMB polarization measurement.
 \end{abstract}
 

 \maketitle
 
 \section{Introduction}
  
  Measurements of the Cosmic Microwave Background, or CMB, are often cited as helping transform cosmology into a precision science. Most of these transformative measurements have been of the non-polarized brightness of the CMB. The CMB has, in addition, a polarized component, measurements of which are beginning to allow us to probe even more deeply into the history and constituents of the Universe. This polarized signal, however, is fainter than the non-polarized part. While building increasingly sensitive instruments to detect this weaker signal, we will inevitably run into dim, new sources of contamination which we could previously ignore, but which will not be negligible in the future. 
 
  Zodiacal light, or Sunlight scattered from dust particles in our Solar System, is often visible with the naked eye. These same interplanetary dust particles also absorb some fraction of the incoming Sunlight, thereby rise in temperature, and then radiate energy at longer wavelengths. With a spectral maximum near 20\,$\mu$m, this interplanetary dust emission (hereafter IPDE) is best observed in the infrared. It began to be characterized in the 20$^\mathrm{th}$ century, and has now been imaged by a number of experiments, including being mapped over the full sky by a few satellite missions. These latter include {\emph{IRAS} }\citep{Neugebauer1984}, which mapped it at 12, 25, 60, and 100\,$\mu$m, and \emph{COBE}/DIRBE, with a wider frequency coverage and better large-scale stability but poorer angular resolution \citep{Kelsall1998}. More recently, \emph{WISE}~\citep{Wright2010}, \emph{Astro-H}~\citep{Pyo2010}, and \emph{Planck}~\citep{Planck2011} have also detected IPDE \citep{Planck2014Zodi} in full-sky maps. A recent review of observations can be found in \citet{Lasue2020}. 
  
\begin{figure}[htp]
   \centering 
   \includegraphics[width=\columnwidth]{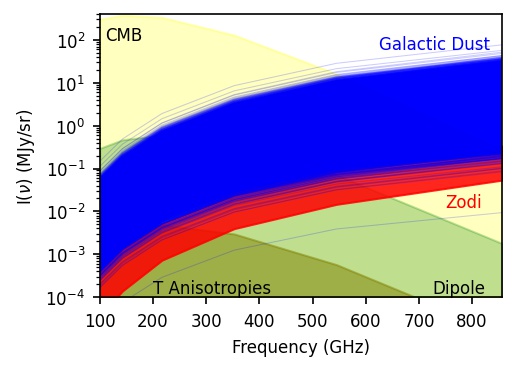}
   \caption{Brightness spectra of IPDE, Galactic dust emission, and CMB components. The total, monopole brightness of the CMB is shown in yellow. The CMB dipole anisotropy is indicated in light green, while a rough idea of the primordial CMB temperature anisotropies are shown in dark green. IPDE and Galactic emissions from Planck are shown in red and blue, respectively. The IPDE estimates cover brightnesses from the Ecliptic plane (the upper part of the shaded red region) to the Ecliptic poles (the lower part of the shaded red region) and are based on the \emph{COBE}/DIRBE model~\protect\citep{Kelsall1998}, with extension to CMB frequencies based on fits to Planck data \citep{Planck2014Zodi}, as described in Section~\protect\ref{sec:ZodiEmissionModel}. Essentially, for each frequency, the IPDE at the ecliptic equator and at the ecliptic pole are calculated with the model, and the values between are filled in red. Galactic emission is represented by an ensemble of light blue lines, each depicting the spectrum of a sky location from the Beyond Planck project's dust model (\protect\url{https://beyondplanck.science/products/files/}). To avoid saturating the graphic, pixel spectra are shown after degrading the maps at each frequency to HEALPix~\protect\citep{Gorski2005} resolution 64, with high transparency.}
   \label{fig:Spectrum}
  \end{figure}

  Though issued from larger grains, in many respects IPDE is similar to thermal emission from grains in our Galaxy -- it is, essentially, thermal emission from dust particles heated by starlight. Given this, it should not be surprising that the frequency dependence of the IPDE brightness is similar to that of Galactic dust. Figure~\ref{fig:Spectrum} shows the range of brightnesses, from \emph{Planck}, expected from both Galactic dust and from IPDE for wavelengths of interest here. 
  
  IPDE, however, is fainter than Galactic emission in total intensity, is in a different magnetic environment, and is significantly less polarized than Galactic dust emission. Though polarization has been detected in scattered Zodiacal light~\citep[see, for example,][]{Pitz1979,Weinberg1980,Leinert1982,Berriman1994}, and though there are mechanisms which might polarize the thermal emission~\citep[for example]{Onaka2000,Hoang2014}, none has been detected to this point. \citet{Siebenmorgen1999} assumed that IPDE was unpolarized in the mid-infrared to set limits of $1\pm0.3$\,\% on instrumental polarization in the \emph{ISO}/CAM instrument. Reversing this argument, one can infer that the mid-infrared polarization fraction of IPDE is at most roughly one percent. 
  
  So while it will always be subdominant to Galactic Dust~\citep{BICEP2018}, since {\emph{Planck}} has already detected IPDE at CMB wavelengths, \emph{LiteBIRD} should detect it as well, with higher significance. While IPDE did not significantly affect the {\emph{Planck}} results, its existence did complicated some aspects of the {\emph{Planck}}/HFI analyses. We must recognize that IPDE models are still very approximate, especially at long wavelengths~\citep{Dikarev2015}, and that the polarization properties of IPDE are still poorly constrained. But the ambitious \emph{LiteBIRD} goals demand that we check that IPDE will not compromise the measurements.
  
  One difficult point in particular is that interplanetary dust is, relatively speaking, close to the Earth. This complicates analyses, because as the Earth orbits around the Sun, the column depth of interplanetary dust through which one observes in a given direction on the distant sky changes depending upon the time of year. As an extreme example, one can imagine observing some distant object which happens to be at an angle close to $180^\circ$ from the Sun (i.e., observing ``away'' from the Sun). If one observes this same point on the sky six months later, the Solar angle will be small (i.e., we'll be observing ``near'' the Sun). The column depth of interplanetary dust through which the observation is done will be larger in the latter case. Additionally, the observation nearer to the Sun will be done though dust which includes some grains that are closer to the Sun and thus hotter than in the first case. When this is combined with changing polarization sensitivity angles, false polarization signals might be inferred. In our toy model, if the first observation were done with a detector sensitive to linear polarization perpendicular to the plane of the ecliptic and the second with a detector sensitive to linear polarization parallel to the plane of the ecliptic, one would incorrectly infer polarization parallel to the plane of the ecliptic.

  To investigate, this work is laid out as follows: the simulations that are used to estimate the size of the effects are described in Section~\ref{sec:Simulations} and the resultant maps and power spectra from these simulations are shown in Section~\ref{sec:Results}. These are discussed in Section~\ref{sec:Discussion} and the paper concludes in Section~\ref{sec:Conclusion}.
  
 \section{Simulations}\label{sec:Simulations}
  
  The simulated observations are based on the {\emph{LiteBIRD}} mission observation strategy described in \citet{Hazumi2020}, but comprise only a single detector at each of 402, 119 and 40\,GHz. The IPDE model used is the so-called \emph{COBE}/DIRBE model~\citep{Kelsall1998}, extended to the longer wavelengths applicable to the CMB using measurements from \emph{Planck} \citep{Planck2014Zodi}. These elements are described below. 
   
  \subsection{\emph{Observation Strategy}}
   
   {\emph{LiteBIRD}} is a CMB mission which will observe in 15 bands between 40 and 402\,GHz. It will observe from orbit around the second Earth-Sun Lagrange point, which is always about 1\% further from the Sun than the Earth, along the Sun-Earth vector. The satellite will ``spin'' around an axis 50 degrees from the focal plane ``boresight'' every 20 minutes, and this spin-axis itself will precess at an angle of 45 degrees about the Sun-Earth vector with a period of 3.2058 hours \citep{Hazumi2020}. The simulation also includes the modulation from a half-wave plate assumed to rotate at a rate of 46, 39, and 61 revolutions per minute at 40, 119\footnote{\emph{LiteBIRD}'s 119-GHz channel actually comprises detectors from both its Low-Frequency Telescope (LFT) and its Mid-Frequency Telescope (MFT). These two telescopes each have half-wave plates which rotate at different frequencies; 46\,rpm for the LFT and 39\,rpm for the MFT. This subtlety does not make a significant difference to the qualitative results presented here, so is ignored.}, and 402\,GHz, respectively. 
   
\begin{figure}[htb]
     \includegraphics[width=\columnwidth]{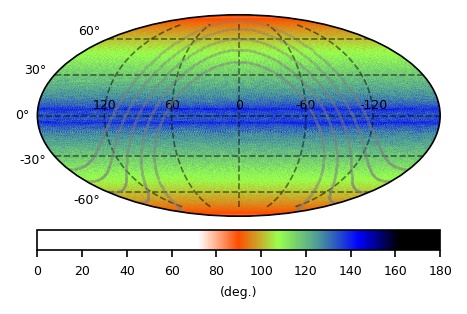}
     \caption{\label{fig:solarAngleAve}Average angle, in degrees, between a given observation direction and the Sun, as seen from the spacecraft, for three years' observations. The minimum average value is about 87 degrees, the median and mean of the map are both about 117 degrees, and the maximum is about 147 degrees. The grey lines mark Galactic latitudes from -20 to 20 degrees.}
   \end{figure}

   As a first step, Figure~\ref{fig:solarAngleAve} shows, for each pixel on the sky in ecliptic coordinates, the average over all \emph{LiteBIRD} samples of the angle between the Sun and the location of the pixel as it is observed from the spacecraft. 
   
   For comparison, we note that \emph{LiteBIRD}'s scanning strategy is better-adapted to minimize the effects of IPDE than that of \emph{Planck}. The density of interplanetary dust is greater closer to the Sun than away from it, and greater towards the ecliptic plane in general than further from it~\citep[see, for example, Figure~4 of][]{Kelsall1998}. One therefore expects larger IPDE signals as as one observes lines-of-sights at angles closer to the Sun. In the ecliptic plane, where IPDE will be greatest, \emph{LiteBIRD} observes at larger elongation angles than did \emph{Planck}, which observed the sky at angle between about 87 and 103$^\circ$ from the Sun. That is, the observation strategy used here looks through lower column depths of interplanetary dust than did \emph{Planck}, on average. As shown in Figure~\ref{fig:solarAngleAve}, in the ecliptic plane, \emph{LiteBIRD} observes at an average angle of about $60^\circ$ from the anti-Solar direction (i.e., about $120^\circ$ from the Sun), while the equivalent angle for \emph{Planck} was about $85^\circ$ (i.e., only about $95^\circ$ from the Sun), though with smaller variations about this average.
   
   \emph{LiteBIRD} will ultimately carry over 4500 detectors across its 15 frequency channels. Using the emission model described below requires integrating over a number of points along each line of sight for each sample of each detector in the simulation. This represents a significant computing undertaking and is beyond the scope of this initial estimation. Therefore, in order to reduce the required computing resources, this work uses the frequencies 40, 119, and 402\,GHz as representative bands. These are, respectively, {\emph{LiteBIRD}}'s longest wavelength band, most sensitive band to CMB, and shortest wavelength band~\citep[see Table~3 of][]{Hazumi2020}. At each of these frequencies, a single detector is simulated for the entire three-year mission. This will obviously gloss over some of the details of how the data from different detectors are combined, but this should not affect the rather broad-brush conclusions in this work. 

   \subsection{\label{sec:ZodiEmissionModel}IPDE Model}
   
    The IPDE model used in this work was developed by the {\emph{COBE} } team in \cite{Kelsall1998}, which is in turn an extension of work done for {\emph{IRAS}}~\citep{Hauser1985}. The model supposes a Diffuse Cloud, three Infrared Dust Bands, and a Circumsolar Ring and ``Trailing Blob''. While \citet{Kelsall1998} used DIRBE data and were thus limited to wavelengths shorter than 240\,GHz, \citet{Fixsen2002} used FIRAS data to extend large-scale observations to longer wavelengths and show that the emissivity of the Diffuse Cloud falls off as $\lambda^{-2}$ above about 150\,$\mu$m. 
    
    With better angular resolution than FIRAS, \citet{Planck2014Zodi} was able to separate the smaller-scale features from the larger-scale Diffuse Cloud and show that the Infrared Dust Bands appear to emit more efficiently at longer wavelengths than does the Diffuse Could. We will see that the Infrared Dust Bands may dominate the IPDE at the longer {\emph{LiteBIRD} } wavelengths. 
    
    The Ring and Blob were not well-detected by Planck, and are therefore neglected here. This is done of necessity, since the long extrapolations done here become untenable with very uncertain parameters. The Ring and Blob, however, are points to keep in mind for \emph{LiteBIRD} in particular, since its observation strategy includes observations with smaller Solar angles than DIRBE and {\emph{Planck}}, and it may therefore observe these two structures more directly than either DIRBE or \emph{Planck}.
    
    While more modern models can now be made \citep[see, for example,][and references therein]{Dikarev2019}, this parameterized \emph{COBE} model should be adequate for giving us an initial understanding of the order of magnitude of the IPDE, and is in a form which is convenient to work with for full-sky CMB maps. In particular, it has also been calibrated at CMB wavelengths by {\emph{Planck} }\citep{Planck2014Zodi}. But once again we underline the approximate nature of the estimates here. 
    
    While the basic elements and geometry of the~\cite{Kelsall1998} model are used here, DIRBE did not observe wavelengths longer than 240\,$\mu$m. \citet{Fixsen2002} added FIRAS data to extend \emph{COBE}'s emissivity measurement of the Diffuse Cloud beyond 240\,$\mu$m, but FIRAS's $7^\circ$ beam prevented them from discussing the Bands, Ring and Blob. In order to make predictions for the longer wavelengths that \emph{LiteBIRD} will observe, therefore, linear fits to the data from Table~3 of \cite{Planck2016HFICalibrationAndMaps} are done in frequency and are interpolated and extrapolated to 402, 119 and 40\,GHz to use for the emissivities of the various IPDE components at \emph{LiteBIRD}'s longer wavelengths. The extrapolation of the Diffuse Cloud emissivity to 40\,GHz, \emph{LiteBIRD}'s longest wavelength, is negative. In this special case, the value is set to zero. Other extrapolation schemes can be used, but these makes no appreciable difference to the results, essentially because the Bands, which \emph{Planck} measured to have much flatter emissivities than the Diffuse Cloud, are the predominate IPDE feature when this model is extrapolated to 40\,GHz. 
    
    \begin{table}[htb]
        \centering
        \caption{40, 119, and 402-GHz emissivities for the Diffuse Cloud and the three IRAS Dust Bands used for the similations.}
        \label{tab:emissivities}
        \begin{tabular}{c|rrr}
            \hline\hline
            Component & 40\,GHz & 119\,GHz & 402\,GHz \\
            \hline
            Cloud  & --0.008 & 0.018 & 0.110 \\
            Band 1 &   1.040 & 1.142 & 1.508 \\
            Band 2 &   0.029 & 0.109 & 0.394 \\
            Band 3 &   0.694 & 0.959 & 1.907 \\
            \hline
        \end{tabular}
        \tablecomments{Emissivity values as interpolated and extrapolated from fits to \emph{Planck}/HFI data \citep{Planck2014Zodi}. In principle, these unitless values should be between 0 and 1. The slightly negative value at 40\,GHz for the Diffuse Cloud is assumedly the result of noise. The values greater than unity arise from the complete degeneracy between the emissivity and the density of particles used in \citet{Kelsall1998}. One could repeat the work in \citet{Planck2014Zodi} with larger densities of particles to ensure these values remain between 0 and 1, but for this work it was more straightforward to simply keep the \citet{Kelsall1998} densities and live with the inconsistency.}
    \end{table}

 \section{Results}\label{sec:Results} 
   
\begin{figure}[htb]
        \centering
        \includegraphics{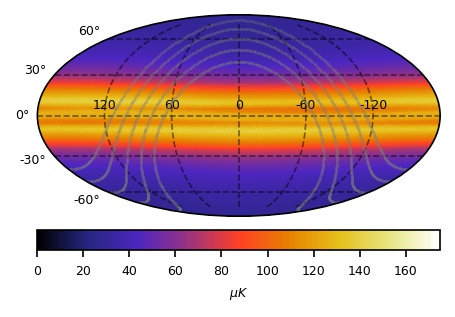}
        \includegraphics{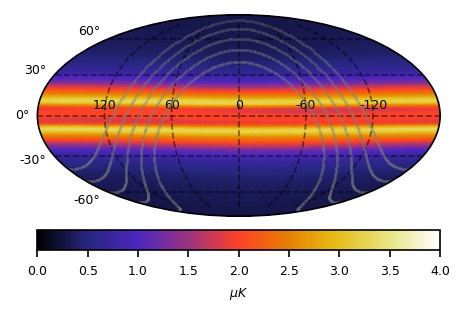}
        \includegraphics{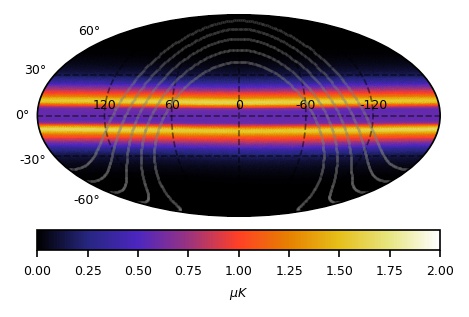}
        \caption{Estimated 402 (top), 119 (middle), and 40-GHz (bottom) temperature (Stokes I) IPDE contributions from a single detector for a mission lasting three years. The maps are in ecliptic coordinates, with the basic structure of the Galactic plane indicated by the series of grey lines, spanning Galactic latitudes from -20 to 20 degrees. }
        \label{fig:TUnFilt}
    \end{figure}

   With the prescription above, unpolarized time-lines of IPDE are simulated for a single, fiducial detector at the center of the focal plane using equation~1 of \citet{Kelsall1998} to integrate along each line-of-sight for each \emph{LiteBIRD} sample taken. These are projected into sky maps using the elementary procedure used for making maps summarized in Section~\ref{sec:MapMaking}. This is repeated for each of 402, 119, and 40\,GHz. The resultant simulated IPDE maps in temperature obtained for three years of observations are shown in Figure~\ref{fig:TUnFilt}. 
    
\begin{figure}[htb]
        \centering
        \includegraphics{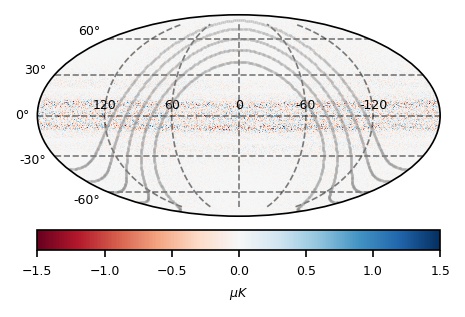}
        \includegraphics{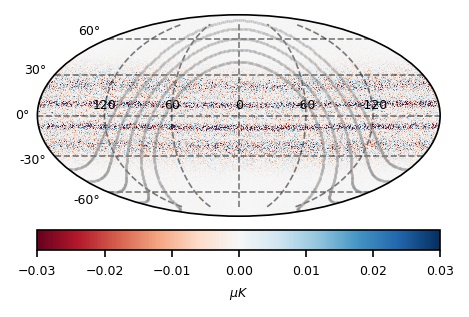}
        \includegraphics{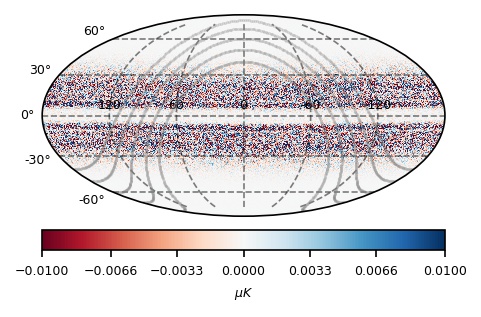}
        \caption{Estimated 402 (top), 119 (middle), and 40-GHz (bottom) Stokes Q IPDE contributions from a single detector for a mission lasting one Solar year. The maps are in ecliptic coordinates, with the basic structure of the Galactic plane indicated by the series of grey lines, spanning Galactic latitudes from -20 to 20 degrees. It should be emphasized that the noise seen here is not due to intrinsic polarization in the IPDE, which is assumed to be zero for these images. It is the result of multiple observations of the distant sky through differing interplanetary dust column depths, the variations in which are ultimately translated as small amounts of polarization by the map-making, which cannot account for this variability.}
        \label{fig:QUnFilt}
    \end{figure}

    While the model emission is assumed to be completely unpolarized, the variability of the column depth of interplanetary dust through which \emph{LiteBIRD} observes could mimic a polarized signal. Imagine a given point on the distant celestial sphere observed once with a given satellite and half-wave plate geometry, and then again, at a later time (usually around six months), with a different set of orientations and angles. Because the column depth of interplanetary dust through which the satellite observes may have changed, the IPDE signal may be different for the two observations, and so a naive observer might infer that the sky is polarized, even when it is not. The apparent change in column depth when observing a given sky direction is significant over timescales of weeks to months (as the satellite orbits the Sun), so the polarization modulation from the fast-spinning half-wave plate helps reduce these effects significantly in Q and U maps. Figure~\ref{fig:QUnFilt} shows plots similar to \ref{fig:TUnFilt}, but for the Stokes Q values. 
    
   The Stokes U maps are qualitatively similar to those of Q in Figure~\ref{fig:QUnFilt}. In these Figures, shown in ecliptic coordinates, one sees the IPDE signal predominantly towards the ecliptic plane, with the thinner signals from the so-called IRAS Bands \citep{Sykes1986} becoming progressively more prominent as one moves from higher to lower frequencies. This is a consequence of the emissivity of the Diffuse Cloud falling more rapidly than that of the IRAS Bands as one observes at longer wavelength (see Table~\ref{tab:emissivities}), and is presumably because the Dust Bands are the result of recent asteroid collisions \citep{Sykes1986} and thus have larger grains than the rest of the cloud, having not yet been ground down with time as has the rest of the Zodiacal Cloud.
    
\begin{figure}[htb]
       \centering
       \includegraphics[width=\columnwidth]{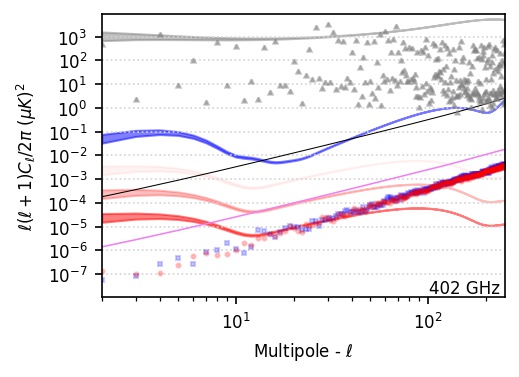}
       \includegraphics[width=\columnwidth]{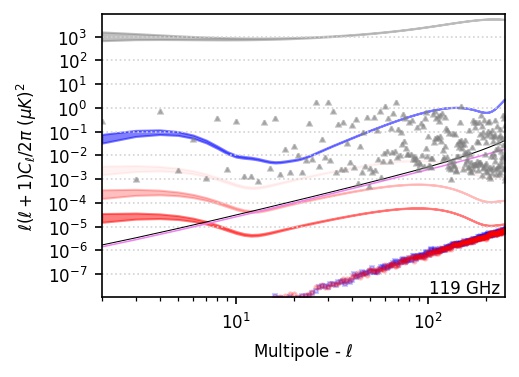}
       \includegraphics[width=\columnwidth]{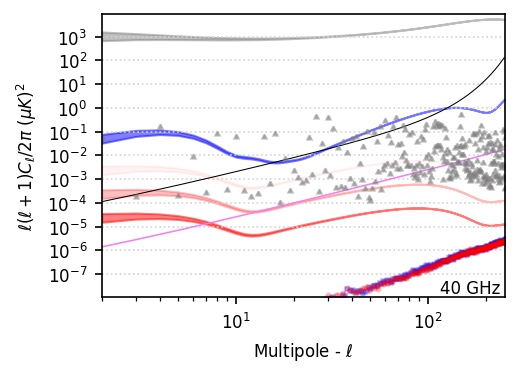}
       \caption{Single detector, noise-free, 402 (top), 119 (middle) and 40-GHz (bottom) temperature (grey triangles), $E$-Mode (blue squares) and $B$-Mode (red circles) power spectra of the maps shown in Figures~\ref{fig:TUnFilt} and \ref{fig:QUnFilt} (plus the analogous U maps). The grey, blue and red curves show the corresponding TT, EE and primordial $B$-mode spectra from gravitational waves with $r=0.1,\ 0.01\ \&\ 0.001$, respectively, in progressively darker red, using parameters from~\protect\citet{Planck2016CosmologicalParameters}. The shaded regions towards the left of each solid curve indicate cosmic variance, the thin black lines represent an estimate of the expected \emph{LiteBIRD} polarization noise levels~\protect\citep{Hazumi2020}, and the violet line represents the spectrum of lensing. The map spectra are made using HEALPix tools~\protect\citep{Gorski2005}.}
       \label{fig:Spectra120}
   \end{figure}

   The TT, EE and BB power spectra for the IPDE maps in Figures~\ref{fig:TUnFilt} and \ref{fig:QUnFilt} (plus the analogous Stokes U maps) are shown in Figure~\ref{fig:Spectra120}, along with models of expected primordial CMB TT, $E$-Mode, and $B$-Mode power spectra for $r$=0.1, 0.01 and 0.001. These spectra are made simply using HEALPix tools~\citep{Gorski2005} without making any sky cuts, since most Galactic cuts don't align with the ecliptic plane and therefore don't significantly affect the IPDE signal. 

   Figure~\ref{fig:Spectra120} also shows the expected \emph{polarization} noise from Table~3 of \citet{Hazumi2020} for each channel as a thin, black line. Note that these curves are for entire \emph{LiteBIRD} channels, after co-addition of all the detectors at the given channel frequency. There may be some subtleties in how the noise in the actual instrument will integrate down. But, since these plots span many orders of magnitude, and since instrumental noise in temperature maps should be the same order of magnitude as instrumental noise in the polarization maps in general, in the order-of-magnitude spirit of these calculations, this one black curve will represent the noise, to be compared with both temperature and polarization power spectra. 

 \section{Discussion}\label{sec:Discussion}
  
   Below, the implications for temperature and polarization measurements are discussed independently.

   \subsection{Temperature}

    Figure~\ref{fig:Spectra120} shows that the IPDE signal is generally smaller than that of the CMB temperature anisotropies across the entire range of \emph{LiteBIRD} frequencies, with the contamination only approaching the level of the CMB near the 402-GHz channel, which will be used as a foreground monitor in any event. In the 119-GHz channel, \emph{LiteBIRD}'s most sensitive to CMB anisotropies, the IPDE is many orders of magnitude below the TT spectrum, and is still lower at 40\,GHz, the long-wavelength end of \emph{LiteBIRD}'s frequency coverage. If one remembers that Galactic foregrounds will need to be accounted for as well (see Figure~\ref{fig:Spectrum}, for example), it is evident that the IPDE is not the primary foreground obstacle for \emph{LiteBIRD}. One might have guessed this, given the results of \emph{Planck} and \emph{WMAP}. 
    
    This does not mean, however, that \emph{LiteBIRD} will not be able to detect the IPDE and that it need not be addressed. Table~2 of \cite{Hazumi2019} quotes noise levels of $47.45, 4.58, \ \mathrm{and}\ 37.42\,\mu$K$\cdot$arc-minute and beam full-width-at-half-maximums of 17.9, 30\footnote{Note that this is a rough average of the resolutions of the two different sets of detectors that contribute to the 119\,GHz channel -- the Low-Frequency Telescope has a full-with-at-half-maximum of 26.3 arc-minutes at 119\,GHz, while the Mid-Frequency Telescope's FWHM is 33.6 arc-minutes.}, and 70.5\,arc-minutes for \emph{LiteBIRD}'s 402, 119, and 40-GHz channels, respectively. The thin, black lines in Figure~\ref{fig:Spectra120} shows the expected sensitivity limits due to instrumental noise for each of our reference channels. When compared with the power spectrum of IPDE in Figure~\ref{fig:Spectra120}, it appears that, at least from the perspective of raw sensitivity, \emph{LiteBIRD} will measure the total power IPDE over its entire frequency range. Even at 40\,GHz, \emph{LiteBIRD}'s longest wavelength and least IPDE-sensitive channel, a number of even-multipole TT power spectrum points, representing the large-angular IPDE structure, are above the \emph{LiteBIRD} noise floor. This indicates that the large-scale structure of IPDE should be detectable, if this imperfect model and its extrapolation in frequency are close to reality. 
    
\begin{figure}[htbp]
        \centering
        \includegraphics[width=\columnwidth]{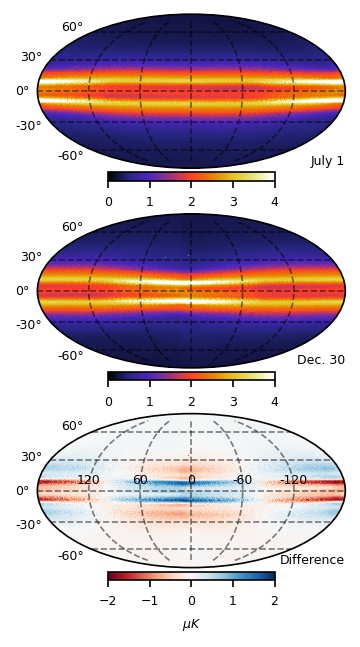}
        \caption{An illustration of map differences that can be used to detect and remove IPDE. The top map is made from six months of simulated 119-GHz \emph{LiteBIRD} IPDE for data taken between July and December. The middle map is made from six subsequent month of data. The bottom map shows the difference of these first two maps. These maps include no noise or other systematics the experiment will also see -- only the IPDE and IPDE differences as seen between successive surveys. Depending upon which frequencies and other detector subsets used to make such differences, these estimated IPDE differences will be hidden somewhat by the associated experimental noise.}
        \label{fig:jackknife}
    \end{figure}

    To make this detection, one will have to remove the primary CMB anisotropies themselves, as well as other more prominent foreground emissions such as Galactic dust and synchrotron emission, as they will be larger than the IPDE signature. But this can effectively be done by making judicious differences of subsets of the data. For example, \emph{LiteBIRD} essentially makes two full surveys of the sky per year. For most locations on the celestial sphere, different column depths of interplanetary dust are seen during successive surveys, even when observing the same direction on the distant sky. This means that differences can be made of maps made of each half-year of data. In these differences, the distant Galactic and cosmological signals will be removed, but a significant fraction of the IPDE will remain in the difference, and can therefore be modeled. This was suggested for {\emph{Planck}} in \citet{Maris2006} and was one of the methods used by the Planck team~\citep{Planck2014Zodi} to address the effects of IPDE in temperature. It is illustrated for \emph{LiteBIRD} in Figure~\ref{fig:jackknife}.
    
\begin{figure}[htbp]
        \centering
        \includegraphics[width=\columnwidth]{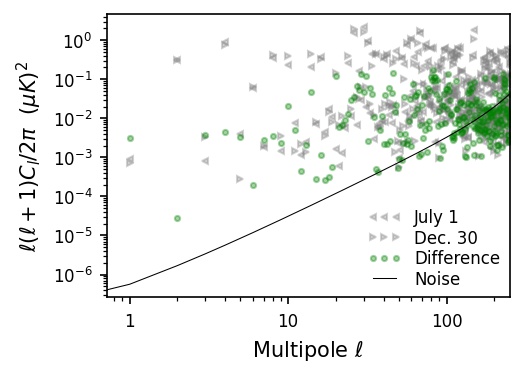}
        \caption{Power spectra of 119-GHz, single-survey IPDE simulations and differences of single-survey IPDE simulations. The power spectrum of a temperature map made from an IPDE simulation of six months of data created with observations beginning July 1, 2029 (i.e., the data shown in the top panel of Figure~\ref{fig:jackknife}) is shown with grey, left-pointing triangles. The power spectrum of a temperature map made from an IPDE simulation of six months of data created with observations beginning December 30, 2029 is shown with grey, right-pointing triangles. The difference is show with the green points. The black line again represents the experimental noise level expected from \protect\citet{Hazumi2020}.}
        \label{fig:JackknifePS}
    \end{figure}

    While it will be difficult to see IPDE in the full temperature maps, once alternating half-year maps are differenced, the IPDE will be clear and still larger than the instrumental noise. This is illustrated in Figure~\ref{fig:JackknifePS}, where the power spectra of both the single-survey maps \emph{and} their difference are shown to be significantly larger than the expected noise at 119\,GHz over a wide range of angular scales of interest. 
    
    It can be modeled or otherwise characterized in order to return to the full maps and remove it before subsequent analyses are done. If necessary, this will need to be done before Galactic component separation and removal, since IPDE will resemble Galactic Dust emission in many ways. 
    
    \subsubsection{Difficulties}
    
     This apparent variability of the IPDE (which, it should be emphasized, is a result in changes in lines-of-sight through the interplanetary dust cloud, not intrinsic variations in the astrophysical source) may also create data reduction confusion. As an example, it would be nice to use the half-year maps in Figure~\ref{fig:jackknife} to try to track systematics such as calibration drifts on six-month time scales. IPDE will, however, complicate this. In Table~\ref{tab:dipole_fits} the result of fits of an offset and a dipole to the first two maps in Figure~\ref{fig:jackknife}, along with the combination of those data (which is quite similar to the middle panel of Figure~\ref{fig:TUnFilt}), are shown.

     \begin{table}[htbp]
      \centering
      \caption{Dipole fits to IPDE simulations presented in Figure~\ref{fig:jackknife}.}
      \begin{tabular}{rccrcrr}
       \hline\hline
       \multicolumn{1}{c}{$\nu$} & Start &  Dur. & Offset & Amp.   & \multicolumn{1}{c}{Lon.}  & \multicolumn{1}{c}{Lat.} \\
       GHz\footnote{The frequency of the simulation in GHz.}   & 
       Month\footnote{The date of the start of the simulation. More precisely, those simulations labeled `July' began on July 1, 2029, while those labeled `Dec.' started December 30, 2029.} & 
       Years\footnote{The length of the simulation in years} & 
       \multicolumn{1}{c}{$\mu$K\footnote{The amplitude of the fitted IPDE offset in $\mu$K.}} & 
       $\mu$K\footnote{The amplitude of the fitted IPDE dipole in $\mu$K.} & 
       deg.\footnote{The ecliptic longitude of the fitted IPDE dipole in degrees.} & 
       deg.\footnote{The ecliptic latitude of the fitted IPDE dipole in degrees.} \\ 
       \hline
       402   & July  & 0.5   & 75.0   & 3.757  & -4.5 & -87.9 \\
       402   & Dec.  & 0.5   & 74.7   & 4.081  & 35.5 &  87.7 \\
       402   & July  & 1.0   & 72.8   & 0.141  & 59.5 & -69.1 \\
       119   & July  & 0.5   &  1.3   & 0.045  &  8.6 & -83.9 \\
       119   & Dec.  & 0.5   &  1.2   & 0.041  &  6.7 &  82.9 \\
       119   & July  & 1.0   &  1.2   & 0.004  & 16.7 & -52.0 \\
        40   & July  & 0.5   &  0.4   & 0.006  & 14.0 & -59.3 \\
        40   & Dec.  & 0.5   &  0.4   & 0.003  & -5.9 & -24.5 \\
        40   & July  & 1.0   &  0.4   & 0.002  &  5.8 & -38.1 \\
        \hline
      \end{tabular}
      \label{tab:dipole_fits}
     \end{table}
     
     There are three lines for each of 402, 119, and 40\,GHz in Table~\ref{tab:dipole_fits}, with the first two lines in each set being the fit to an offset and a dipole to six months of data -- the data shown in the first two images of Figure~\ref{fig:jackknife} for 119\,GHz, for example. The third line in each of these three sets shows the same fit to the combination of these data in a one-year map. It should be first noted that while the Dipole amplitude found for the full-year data are quite small, the intermediate fits on six-month subsets are much larger. 
     
     \citet{Delouis2021} quote a staggeringly small 40\,nK uncertainty on the amplitude of the ~3.4\,mK dipole amplitude at 100 and 143\,GHz, and 710\,nK at 353\,GHz with \emph{Planck}, which is larger than the IPDE dipole amplitude found on six-month timescales (though not yearly time scales). Therefore, if no corrections for the IPDE are made, its apparent time variability will even limit the measurement of the stability of the calibration on six-month time scales with the Dipole. 
    
     While tracking calibration drift can be done on \emph{yearly} temperature map differences, it is more difficult for some other systematic effects. Far sidelobes, for example, are notoriously difficult to characterize. They too appear as faint, slowly varying, large-scale features in the maps, they too will show differences in half-year maps, but not in yearly maps, and they therefore may need to be separated from IPDE with care. So, while smaller than the CMB cosmic variance levels in those channels sensitive to the CMB, and while large-scale, total power measurements are no longer the primary target of future CMB experiments, this signal will be detected and must be fitted, filtered, or otherwise removed or accounted for.
   
   \subsection{\label{sec:Polarization}Polarization}
   
    Here, the IPDE is initially assumed to be unpolarized, but there will sometimes be certain combinations of data taken at different times with different polarization and half-wave plate orientations which might mimic a polarization signal. Figure~\ref{fig:Spectra120} also shows that the noise induced in \emph{LiteBIRD} polarization maps from IPDE observed through varying column depths of interplanetary dust is negligible compared to instrumental noise if the IPDE is inherently non-polarized. This is good news as far as \emph{LiteBIRD}'s quest for CMB $B$ Modes is concerned. 
    
    So, similar to the case of temperature, the largest effects from IPDE for polarization may once again be in complications in data analysis pipeline. The apparent dependence of the signal on time and on the orientation of the detectors will make, for example, null tests based on the differences between different detectors more delicate to design and interpret at the faintest levels. Naive tests made by differencing the data from a detector with one orientation from that with another will fail due to IPDE, and might lead one to incorrectly assume that there is something wrong with the data when there is not. 
    
    \subsubsection{Intrinsic Polarization}
    
    There are important caveats to the assertion that the IPDE is unpolarized: in preparing this work, the best limits found on the polarization fraction of thermal IPDE (as opposed to scattered light) are about 1\%~\citep[from \emph{ISO}/CAM;][]{Siebenmorgen1999}. It should also be noted that these \emph{ISO}/CAM observations were designed not to measure the IPDE, but simply to use IPDE to set limits on the \emph{ISO}/CAM instrumental polarization. They consist of a few hours of observations taken on a single day (February 8, 1997), observing towards a single location on the sky (equatorial lon./lat.=01h21m50.0s/+12d20m00s)\footnote{See the \emph{ISO} Observation Log at \url{https://heasarc.gsfc.nasa.gov/W3Browse/iso/isolog.html}}. Since this 1\% is also roughly the polarization fraction that can, at least in principle, be created via mechanisms such as those presented in \citet{Onaka2000} and \citet{Hoang2014}, it is a convenient target to test.
    
\begin{figure*}[htb]
        \centering
        \includegraphics{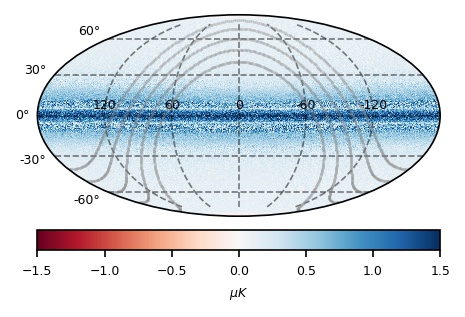}
        \includegraphics{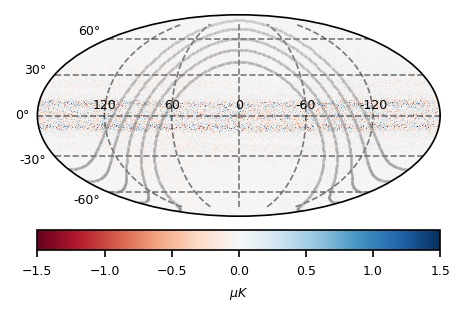}
        \includegraphics{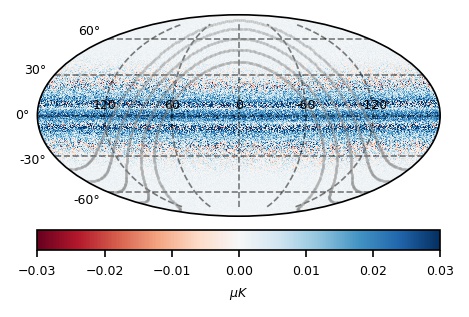}
        \includegraphics{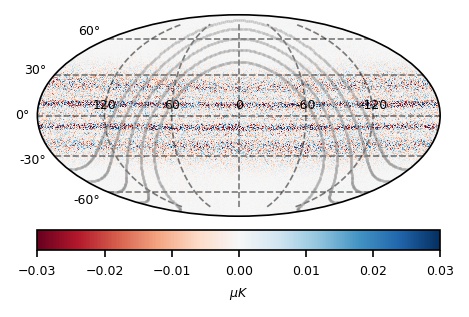}
        \includegraphics{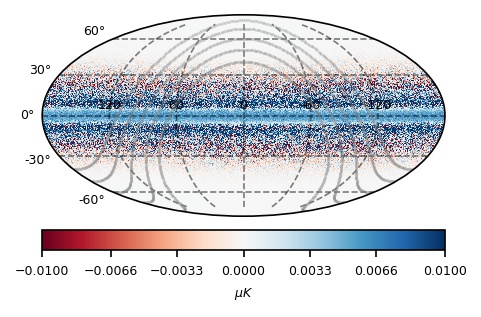}
        \includegraphics{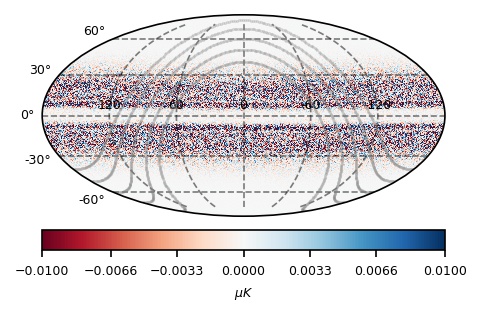}
        \caption{Similar to Figure~\ref{fig:QUnFilt}, but assuming that the IPDE is 1\% polarized. Estimated 402 (top), 119 (middle), and 40-GHz (bottom) Stokes Q (left) and U (right) IPDE contributions from a single detector for a mission lasting one Solar year. The maps are in ecliptic coordinates, with the basic structure of the Galactic plane indicated by the series of grey lines, spanning Galactic latitudes from -20 to 20 degrees. }
        \label{fig:QUnFiltPolar}
    \end{figure*}

    To this end, the simulations described previously have been redone with a toy model, assuming that the IPDE is 1\%-polarized. Lacking any other information, the polarization direction is simply assumed to be perpendicular to the plane defined by the Sun--satellite line, and the line-of-sight. This is what was found in \citet{Berriman1994} for scattered polarization and is broadly consistent with models such as that from \citet{Onaka2000}. While one would also expect some dependence on the angle between these two lines, for example, this is ignored here -- the polarization amplitude is assumed to be 1\% of the unpolarized amplitude, regardless of observation geometry. The Stokes Q and U maps obtained with these changes are shown in Figure~\ref{fig:QUnFiltPolar}. 
    
    \vspace{0.25cm} 
    
\begin{figure}[htb]
       \centering
       \includegraphics[width=\columnwidth]{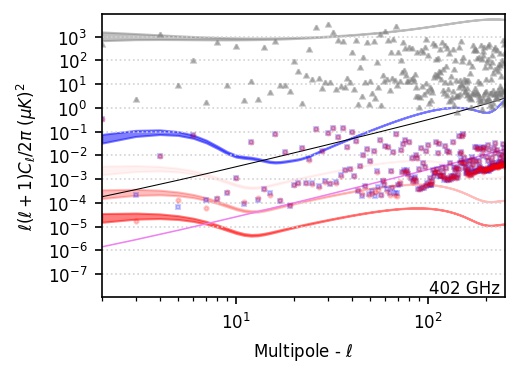}
       \includegraphics[width=\columnwidth]{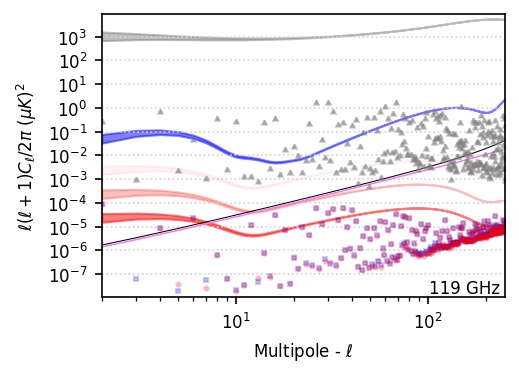}
       \includegraphics[width=\columnwidth]{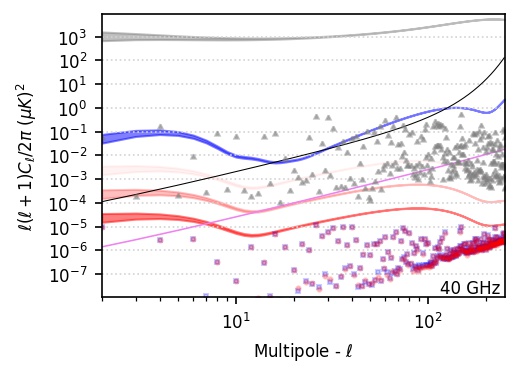}
       \caption{Similar to Figure~\ref{fig:Spectra120}, but with 1\% polarization for the emission, assumed to be in the scan direction. Single detector,  noise-free, 402 (top), 119 (middle) and 40-GHz (bottom) temperature (grey triangles), $E$-Mode (blue squares) and $B$-Mode (red circles) power spectra of the maps shown in Figures~\ref{fig:TUnFilt} and \ref{fig:QUnFiltPolar}. The grey, blue and red curves show the corresponding TT,  EE and primordial $B$-mode spectra from gravitational waves with $r$=0.1, 0.01 \& 0.001, respectively, in progressively darker red, using parameters from~\protect\citet{Planck2016CosmologicalParameters}. The shaded regions towards the left of each solid curve indicate cosmic variance, the thin black lines represent an estimate of the expected \emph{LiteBIRD} polarization noise levels~\protect\citep{Hazumi2020}, and the violet line represents the spectrum of lensing. The spectra are made using HEALPix tools~\protect\citep{Gorski2005}.}
       \label{fig:Spectra1percent0}
    \end{figure}

    While the high-spatial-frequency noise seen in Figure~\ref{fig:QUnFilt} remains, if the IPDE is polarized, Figure~\ref{fig:QUnFiltPolar} shows that there will be large-scale structure in the Stokes Q maps, while the U map remains similar to that without IPDE polarization. The relative lack of U versus Q signal arises from the geometry of the observations and our fiat that the polarization will always be perpendicular to the plane containing the Sun-Earth vector and the observation line of sight. Near the ecliptic plane, which contains the bulk of the IPDE, this constrains the polarization to be along the ecliptic z-axis -- hence mostly (negative) Q Stokes signal. The power spectra resulting from these maps are shown in Figure~\ref{fig:Spectra1percent0}.
    
    A first take-away from Figure~\ref{fig:Spectra1percent0} is that \emph{LiteBIRD} should, in fact, be able to improve upon current limits on the polarization fraction of IPDE. The top panel shows, for example, that the low, even multipoles of the IPDE should be detected with high significance at 402\,GHz. In fact, if the model used here is reasonable, the signal should be detectable up to multipoles of order 30 or so. On a less positive note, while IPDE polarization will not be detectable at the lowest \emph{LiteBIRD} frequencies, if the actual polarization fraction of IPDE is actually close to the current 1\% limit, IPDE could actually affect the CMB $B$-Mode power spectrum in the 119-GHz and neighboring channels. In this case, as was the case with temperature, the signal will need to be masked, modeled and/or similarly removed. If one were to add a $15^\circ$ ecliptic plane mask in addition to a fairly typical $30^\circ$ Galactic plane mask, the usable sky fraction goes from ~50\% of the sky after the Galactic cut to ~34\% after the Galactic plus Ecliptic rejections -- that is, an additional Ecliptic cut will remove over 30\% of the remaining sky. 
   
 \section{Conclusion}\label{sec:Conclusion}
  
   For analyses that need to approach \emph{LiteBIRD}'s instrumental noise limits, IPDE will need to be removed from the data, as was the case for the \emph{Planck} mission and as has been the case for many infrared measurements for some time. In temperature, however, Figure~\ref{fig:Spectra120} shows that CMB power spectrum contamination will be at the level of a percent of the cosmic variance of the CMB or even lower. 
   
   Given that \emph{LiteBIRD} observes at wavelengths a factor of 50 away from the peak of IPDE, it is a testament to the exquisite polarization sensitivity target that \emph{LiteBIRD} should be able to set new limits on the IPDE. On the other hand, if the intrinsic polarization is close to the current limits, the induced $B$-Mode signal will be roughly the same size as the \emph{LiteBIRD} CMB $B$-Mode target at 119\,GHz, and this will therefore present an additional wrinkle added to Galactic foreground removal. This should be kept in mind as \emph{LiteBIRD} data reduction and analysis projects progress. 
   
   Before finishing, it's worth recalling that IPDE models are fairly uncertain at these ``long wavelengths'' of interest to CMB experimenters. \cite{Dikarev2015} have compared the \cite{Kelsall1998} model with two other, more recent models and found significant differences. While their \cite{Divine1993} and \cite{Dikarev2005} models have less ``structure'' in them, the absolute level of the contamination is, in fact, estimated to be higher than that in the \cite{Kelsall1998} model, and may, therefore, show more contamination in CMB maps \citep[see also][]{Jorgensen2021}. With this caveat, a more thorough comparison is deferred to a later work.
   
   Finally, we note that Figure~\ref{fig:Spectrum} reinforces the notion that Galactic dust emission will almost always dominate over IPDE. To first order, IPDE is less bright than Galactic dust emission. But the IPDE brightness spectrum, as noted previously, is similar to that of Galactic dust emission. In fact, it is similar enough that this may make separation of the two difficult. While it may be a coincidence that the brighter regions of IPDE are similar in brightness to the dimmer regions of Galactic emission, it may also indicate some confusion between the two. In addition, the interplanetary dust cloud's proximity also leads to an apparent brightness variation, and it is here that IPDE may be at its most annoying. This apparent variability will limit somewhat tests and analyses and make removing Galactic dust to below the level of the IPDE more difficult. But given the differences in amplitude and polarization fraction, IPDE should ultimately not be as difficult to address as Galactic emissions. 
   
 \begin{acknowledgements}
  Nils Odegard and Janet Weiland signaled problems with earlier versions of the code used here as part of their work on \citet{Odegard2019}, and Martin Bucher, Josquin Errard, and Marc Sauvage provided useful elements, discussions and comments. The bulk of the computing for this work was done at the IN2P3's Computing Center (\url{https://cc.in2p3.fr}). Mathieu Remazeilles acknowledges funding support from the ERC Consolidator Grant CMBSPEC (No. 725456) under the European Union's Horizon 2020 research and innovation program. 
  
  This work is supported in \textbf{Japan} by ISAS/JAXA for Pre-Phase A2 studies, by the acceleration program of JAXA research and development directorate, by the World Premier International Research Center Initiative (WPI) of MEXT, by the JSPS Core-to-Core Program of A. Advanced Research Networks, and by JSPS KAKENHI Grant Numbers JP15H05891, JP17H01115, and JP17H01125. The \textbf{Italian} \emph{LiteBIRD} phase A contribution is supported by the Italian Space Agency (ASI Grants No. 2020-9-HH.0 and 2016-24-H.1-2018), the National Institute for Nuclear Physics (INFN) and the National Institute for Astrophysics (INAF). The \textbf{French} \emph{LiteBIRD} phase A contribution is supported by the Centre National d’Etudes Spatiale (CNES), by the Centre National de la Recherche Scientifique (CNRS), and by the Commissariat à l’Energie Atomique (CEA). The \textbf{Canadian} contribution is supported by the Canadian Space Agency. The \textbf{US} contribution is supported by NASA grant no. 80NSSC18K0132. \textbf{Norwegian} participation in \emph{LiteBIRD} is supported by the Research Council of Norway (Grant No. 263011). The \textbf{Spanish} \emph{LiteBIRD} phase A contribution is supported by the Spanish Agencia Estatal de Investigación (AEI), project refs. PID2019-110610RB-C21 and AYA2017-84185-P. Funds that support the \textbf{Swedish} contributions come from the Swedish National Space Agency (SNSA/Rymdstyrelsen) and the Swedish Research Council (Reg. no. 2019-03959). The \textbf{German} participation in \emph{LiteBIRD} is supported in part by the Excellence Cluster ORIGINS, which is funded by the Deutsche Forschungsgemeinschaft (DFG, German Research Foundation) under Germany’s Excellence Strategy (Grant No. EXC-2094 - 390783311). This research used resources of the Central Computing System owned and operated by the Computing Research Center at KEK, as well as resources of the National Energy Research Scientific Computing Center, a DOE Office of Science User Facility supported by the Office of Science of the U.S. Department of Energy.
 \end{acknowledgements}
 
 \bibliographystyle{aasjournal}
 \bibliography{KenGanga}
 
 \appendix
 
 \section{Map-Making}
   \label{sec:MapMaking}
   
   Assuming white (or no) noise, we can make temperature and polarization maps using the equation
   \begin{equation}
    \mathbf{m}
    =
    \left(\mathbf{A}^T\mathbf{A}\right)^{-1}\mathbf{A}^T\mathbf{d}. 
    \label{eq:MapMakingEquation}     
   \end{equation}
   Here, $\mathbf{d}$ is the vector of length $n_\mathrm{samples}$ with our time-line of IPDE values for each 19\,Hz sample, $i$. That is, 
   $$
    \mathbf{d}
    =
    \left(
     \begin{array}{c}
      d_1 \\
      d_2 \\
      d_3 \\
      \vdots \\
      d_{n_\mathrm{samples}}
     \end{array}
    \right).
   $$
   
   The compact notation in equation~\ref{eq:MapMakingEquation} hides the fact that $\mathbf{m}$ is actually a set of three maps, one for each of the Stokes parameters T, Q, and U (we avoid `I', which is being used as a sample index in lower case). That is,
   $$
    m_{s,p}
    =
    \left(
     \begin{array}{ccc}
      t_1                   & q_1                   & u_1 \\
      t_2                   & q_2                   & u_2 \\
      t_3                   & q_3                   & u_3 \\
      \vdots                & \vdots                & \vdots \\
      t_{n_\mathrm{pixels}} & q_{n_\mathrm{pixels}} & u_{n_\mathrm{pixels}} \\
     \end{array}
    \right),
   $$
   where $s$ is an index over the three Stokes parameters and $p$ is an index over all map pixels. The so-called ``pointing matrix'', $\mathbf{A}$, is defined such that
   $$
    A_{i,1,p} 
    = 
    \left\{ 
     \begin{array}{cl}
      1 & \mathrm{if\ sample}\ i\ \mathrm{lands\ in\ pixel}\ p \\
      0 & \mathrm{otherwise}
     \end{array}
    \right.,
   $$
   
   $$
    A_{i,2,p} 
    = 
    \left\{ 
     \begin{array}{cl}
      \cos\left(2\theta_i\right) & \mathrm{if\ sample}\ i\ \mathrm{lands\ in\ pixel}\ p \\
      0 & \mathrm{otherwise}
     \end{array}
    \right.,
   $$
   and
   $$
    A_{i,3,p} 
    = 
    \left\{ 
     \begin{array}{cl}
      \sin\left(2\theta_i\right) & \mathrm{if\ sample}\ i\ \mathrm{lands\ in\ pixel}\ p \\
      0 & \mathrm{otherwise}
     \end{array}
    \right..
   $$
   This means that for pixel $p$, 
   \begin{eqnarray}
    \mathbf{A}^T\mathbf{A}
    & = &
    \sum_{s_p}
    \left(
     \begin{array}{ccc}
      1                  & \cos 2\theta_{s_p}                   & \sin 2\theta_{s_p} \\
      \cos 2\theta_{s_p} & \cos^2 2\theta_{s_p}                 & \cos 2\theta_{s_p} \sin 2\theta_{s_p} \\
      \sin 2\theta_{s_p} & \cos 2\theta_{s_p}\sin 2\theta_{s_p} & \sin^22\theta_{s_p}
     \end{array}
    \right),
    \nonumber
   \end{eqnarray}
   \\
   \noindent
   where $s_p$ is the set of samples $s$ which fall in pixel $p$. This depends only on the scanning and sampling patterns. 
  
\end{document}